\documentclass[epj]{svjour}
\usepackage{epsfig} 
\usepackage{psfig} 
\usepackage{times}
\begin{document}
\title{%\hfill{\tiny FZJ-IKP-TH-2005-12, HISKP-TH-05/09}\\
The {\boldmath $K^-\alpha$} scattering length and  the reaction
{\boldmath $dd{\to}\alpha{K^+K^-}$}}
\author{V.~Yu.~Grishina\inst{1}, L.A.~Kondratyuk\inst{2}, 
A.~Sibirtsev\inst{3}, M.~B\"uscher\inst{4}, S.~Krewald\inst{4},
U.-G.~Mei{\ss}ner\inst{3,4} and F.~P.~Sassen\inst{3}} 
\institute{Institute for Nuclear Research,
60th October Anniversary Prospect 7A, 117312 Moscow, Russia \and
Institute of Theoretical and Experimental Physics, B.\
Cheremushkinskaya 25, 117259 Moscow, Russia \and 
Helmholtz-Institut f\"ur Strahlen- und Kernphysik (Theorie), 
Universit\"at Bonn, Nu\ss allee 14-16, D-53115 Bonn, Germany \and
Institut f\"ur Kernphysik, Forschungszentrum J\"ulich,
D-52425 J\"ulich, Germany }
\date{Received: date / Revised version: date}

\abstract {We present predictions for the the $K^- \alpha$
scattering length  obtained within the framework of the multiple 
scattering approach. 
Evaluating the pole position of the $K^-\alpha$ scattering amplitude
within the  zero range approximation, we find a loosely bound
$K^- \alpha$ state with a  binding energy 
of $E_R{=}-2{\ldots}-7$~MeV and a width $\Gamma_R{=}11{\ldots}18$~MeV. 
We propose to measure the $K^-\alpha$ scattering length through
the final state interaction between the $\alpha$ and $K^-$-meson 
produced in the reaction  $dd{\to}\alpha{K^+K^-}$. 
It is found  that  the $K^-\alpha$ invariant mass distribution from 
this reaction at energies near the threshold provides a new tool
to determine the $s$-wave $K^-\alpha$ scattering length.
}


\authorrunning{V. Yu. Grishina et al.} \titlerunning{
$K^-\alpha$ scattering length}

\maketitle

\section{Introduction}

Low-energy ${\bar K}N$ and ${\bar K}A$
interactions have gained substantial interest during the last 
two decades. It is known from the time-honored Martin analysis~\cite{Martin}
that the isoscalar $s$-wave $K^-N$ scattering length is large and repulsive, 
Re$a_0{=}{-}1.7$~fm, while
the isovector length is moderately attractive,  Re$a_1{=}0.37$~fm.  
It is clear that such a strong repulsion in the 
${\bar K}N$ isoscalar channel leads
also to a repulsion in the low-energy $K^-p$ system, since
Re$a_{K^- p}{=}0.5\mathrm{Re}(a_0{+}a_1)$=$-$0.74~fm.
It should be noted that Conboy's analysis~\cite{Conboy} of 
low energy ${\bar K}N$ data gives a  solution with Re$a_0$=$-$1.03~fm
and Re$a_1$=0.94~fm, that also results in  repulsion in the 
$K^-p$ channel, but with substantially smaller strength,
Re$a_{K^- p}$=$-$0.05 fm. Data from KEK 
show that the energy shift of the 1$s$
level of kaonic hydrogen is repulsive~\cite{Ito}.
Very recent results for kaonic hydrogen from the DEAR experiment
\cite{Guaraldo1} also indicate a repulsive energy shift.  
However, the consistency of the bound state with the scattering
data can be questioned, as first pointed out in Ref.~\cite{MRR}.

Nevertheless, it is possible that the actual $K^-p$ interaction is
attractive if the isoscalar $\Lambda(1405)$ resonance is a bound state
of ${\bar K}N$ system~\cite{Dalitz,Weise1}. A fundamental reason
for such a scenario is provided by the leading order term in the chiral
expansion for the $K^-N$ amplitude which is attractive. 
New developments in the analysis of the ${\bar K}N$ interaction based on 
chiral Lagrangians can be found in Refs.\cite{Weise,Oset,Oller,LuKo}.
These results provide further support for the description
of the $\Lambda$(1405) as a meson-baryon bound state. More recently,
it has even been argued that there are indeed two poles in the complex
plane in the vicinity of the nominal $\Lambda$(1405) pole~\cite{Jido}.
For recent evidence to support this scenario, see e.g.~\cite{OsetJ}.
A different view seems to be taken in Ref.\cite{BNW}.

Such a non-trivial dynamics of the ${\bar K}N$ interaction leads to
very interesting in-medium phenomena in interactions of anti-kaons 
with finite nuclei as well as with dense nuclear matter, 
including neutron stars, see e.g.
Refs.~\cite{Sibirtsev1,Lutz,Sibirtsev2,Ramos,Heiselberg,Cieply}.

Recently, exotic few-body nuclear systems involving the $\bar K$-meson as
a constituent were studied by Akaishi and Yamazaki~\cite{Akaishi}.
They proposed a phenomenological ${\bar K}N$ potential model, which 
reproduces the $K^-p$ and $K^-n$ scattering lengths from the Martin
analysis~\cite{Martin}, the kaonic hydrogen atom data from 
KEK~\cite{Ito,Iwasaki} and the mass and width of the 
$\Lambda$(1405) resonance. The ${\bar K}N$ interaction in this model
is characterized by a strong $I{=}0$ attraction, which allows the 
few-body systems to form dense nuclear objects. As a result, the nuclear 
ground states of a $K^-$ in $(pp)$, $^3$He, $^4$He and $^8$Be were
predicted to be discrete states with binding energies of 48, 108, 86
and 113 MeV and widths of 61, 20, 34 and 38 MeV, respectively. 
More recent work on this subject can be found e.g. 
in Refs.~\cite{Dote1,Dote2}.

Furthermore, very recently a strange tribaryon $S^0(3115)$ was detected
in the interaction of stopped $K^-$-mesons with 
$^4$He~\cite{Suzuki}. Its width was found to be less 
than 21 MeV. In principle, this state may  be interpreted
as a candidate of a deeply bound state $({\bar K}NNN)^{Z{=}0}$ with 
$I{=}1,I_3{=}{-}1$. However, the observed tribaryon $S^0(3115)$ is 
about 100 MeV
lighter than the predicted mass. Moreover, in the experiment 
an isospin~1 state was detected at a position where no peak was predicted.
Further searches for bound kaonic nuclear states as well as new 
data on the interactions of $\bar K$-mesons with lightest nuclei are thus
of  great importance.

Up to now the $s$-wave $K^-\alpha$ scattering 
length, which we denote as $A(K^- \alpha)$, has not been measured and 
relevant theoretical calculations  have not yet
been done. In this paper we present a first calculation 
of $A(K^-\alpha)$ within the framework of the multiple scattering 
approach (MSA). 

We investigate the pole position of the $K^-\alpha$ scattering amplitude 
within the zero range approximation (ZRA) in order to find out whether
the formation of a bound state in $\bar K \alpha$ system is possible.
Furthermore, we discuss the possibility to measure the ${\bar K}\alpha$
scattering length through  the ${\bar K}\alpha$  final state 
interaction (FSI). Recently it was proposed to measure the reaction
$dd{\to}\alpha{K^+ K^-}$ near the threshold at 
COSY-J\"ulich~\cite{Buescher02}. 
We apply our approach to calculate the $K^- \alpha$
FSI effect in this reaction and demonstrate that the 
$K^-\alpha$ invariant mass distribution is sensitive enough to 
the $K^-\alpha$ FSI  and may be used for a
determination of the $s$-wave $K^-\alpha$ scattering length. 

Our paper is organized as follows: In Sect.~2 we calculate
the $K^-\alpha $ scattering length within the MSA and determine the 
pole position of the amplitude in the zero range approximation.
In Sect.~3 an analysis of the FSI 
in the reaction  $dd{\to}\alpha{K^+ K^-}$ is considered.
Our conclusions are given in Sect.~4.

\section{The {\boldmath $K^-\alpha$} scattering length}
      
\subsection{Multiple scattering formalism}
To calculate the $s$-wave $K^-\alpha$ scattering length as well as
the FSI enhancement factor, we use the Foldy--Brueckner adiabatic approach
based on the multiple scattering (MS) formalism~\cite{Goldberger}. 
Note that this method has already been used for the calculation of the 
enhancement factor in the reactions 
$pd {\to} ^3$He$\eta$ \cite{Faldt2}, $pn {\to} d\eta$ \cite{Grishina1} and 
$pp {\to} d\bar{K^0}K^+$ \cite{Grishina_a0_04}.

In the Foldy--Brueckner adiabatic approach, the continuum 
$K^-\alpha$ wave function, which is defined at fixed coordinates 
of the four nucleons in $^4$He,  can be written as the sum of the incident
plane wave of the kaon and waves emerging from the four fixed
scattering centers. Keeping only the $s$-wave contribution, we can
express the total wave function $\Psi_k$ through the $j$-channel wave 
functions ${\psi}_j({\mathbf r}_j)$  in the following way
\begin{eqnarray}
\Psi_k({\mathbf r}_{K^{-}}{;}{\mathbf r}_1,{\mathbf r}_2,
{\mathbf r}_3,{\mathbf r}_4){=} 
{\mathrm e}^{i{\mathbf k}{\cdot}
\mathbf {r}_{K^-}}\!\!
{+} \!\!\sum_{j=1}^{4}  \!t_{K^-N_j} \frac{{\mathbf e}^{i k R_j}}{R_j}
\ {\psi}_j({\mathbf r}_j), \label{fixcenter4}
\end{eqnarray}
where $R_j{=}\left|\mathbf{r}_{K^-}{-}\mathbf{r}_j\right|$ and  the
t-matrix, $t_{K^-N_j}$, is related to the elastic scattering amplitude
$f_{K^- N}$ via~\cite{Grishina1,Grishina_a0_04}
\begin{equation}
t_{K^- N}(k_{K^- N}) =
(1+\frac{m_{K^-}}{m})\, f_{K^- N}(k_{K^- N}),
\end{equation}
with $m \,(m_{K^-})$ the nucleon (charged kaon) mass, 
and $k_{\bar{K}N}$ is the modulus of the relative $\bar{K}N$ momentum.
Note that we use the unitarized scattering length approximation
for the latter, {\it i.e.}
\begin{equation}
f^{I}_{\bar{K} N}(k_{\bar{K} N})=
\left[(a^{I}_{\bar{K} N})^{-1} - ik_{\bar{K} N}\right]^{-1},
\end{equation}
where $I$ is the isospin of the $\bar{K}N$ system. For each
scattering center $j$ an effective wave ${\psi}_j(\mathbf {r}_j)$
is defined as the sum of the incident plane wave and the waves
scattered from the three other centers
\begin{eqnarray}
{\psi}_j({\mathbf r}_j)= {\mathrm e}^{i{\mathbf k} \cdot
{\mathbf r}_j}+ \sum_{l \neq j} t_{K^- N_l} \frac{{\mathbf e}^{i k
R_{jl}}}{R_{jl}} \ {\psi}_l({\mathbf r}_l)\ ,
\label{fixcenterwave}
\end{eqnarray}
where $R_{jl}{=}\left |\mathbf{r}_l{-}\mathbf{r}_j \right|$.
Therefore, the channel wave functions ${\psi}_j({\mathbf r}_j)$ can
be found by solving the system of the four linear
equations~(\ref{fixcenterwave}).

To obtain the FSI factor we calculate the total wave 
function $\Psi_k$ given by Eq.~(\ref{fixcenter4}) at 
$\mathbf{r}_{K^-}{=}\sum_{j=1}^4\mathbf{r}_j{=}0$ and average
it over the coordinates of the nucleons ${\mathbf{r}}_j$ in $^4$He.
Thus the FSI enhancement factor is~\cite{Goldberger}
\begin{eqnarray}
\lambda^{\mathrm{MS}}(k_{K^- \alpha}) {=}
  \left| \left\langle   \Psi_{q_{K^-}^{\mathrm{lab}}}
(\mathbf{r}_{K^{-}}\!{=}\!\!\sum_{j{=}1}^4\!
  \mathbf{r}_j{=}0;\mathbf{r}_1,\mathbf{r}_2,\mathbf{r}_3,
\mathbf{r}_4)\right\rangle \right|^2\!\!. 
\label{enhancement}
\end{eqnarray}
For the nuclear density function we use the factorized form
\begin{eqnarray}
&&\left|\mathrm
{\Phi}(\mathbf{r}_1,\mathbf{r}_2,\mathbf{r}_3,\mathbf{r}_4)
\right|^2= \prod_{j=1}^4 \rho_{j}({\mathbf {r}}_j) \ , \label{he4}
\end{eqnarray}
where the single nucleon density is taken in  Gaussian form as
\begin{eqnarray}
\rho (\mathbf{r})= \frac{1}{(\pi \, R^2)^{3/2}} \
\mathrm{e}^{-r^2/R^2} \ , 
\label{radiushe4}
\end{eqnarray}
with $R^2/4{=}0.62$~fm$^{2}$. Note that the independent
particle model formulated by Eqs.~(\ref{he4}-\ref{radiushe4}) 
provides a rather good description of the $^4$He  electromagnetic 
form factor up to momentum transfer 
$\mathbf{q}^2{=}8$~fm${}^{-2}$~\cite{Boitsov}.

The integration in Eq.~(\ref{enhancement}) over the
nucleon coordinates~$\mathbf{r}_j$ was performed
using the Monte-Carlo method. This approach provides us with a
possibility to include all  configurations of the nucleons
in ${}^4 \mathrm{He}$. Within this method we can also take into
account in Eq.~(\ref{fixcenter4}) the dependence of the $t_{K^- N_j}$
amplitude on the type of nucleonic scatterer, {\it i.e.} proton or neutron.
Note that the simple version of the multiple scattering approach
used in Ref.~\cite{Wycech} can be applied only to the case of
identical scatterers.

The s-wave $K^-\alpha$ scattering length can be derived from the
asymptotic expansion of Eq.~(\ref{fixcenter4}) at 
$r_{K^-}{\to}\infty$ and it is 
\begin{eqnarray}
A(K^- \alpha)=\frac{m_{\alpha}}{m_{\alpha}+m_{K^{-}}}
\left.\left\langle\sum_{j=1}^{4} t_{K^- N}  \
{\psi}_j({\mathbf r}_j) \label{akhe4}\right\rangle
\right|_{\sum_{j=1}^4
  \mathbf{r}_j =0}~,
\end{eqnarray}
with $m_\alpha$ the $\alpha$-particle mass.
Here the procedure of averaging over the coordinates of the nucleons
is similar to Eq.~(\ref{enhancement}). 

\subsection{S-wave scattering length and the pole position
of the amplitude in the zero range approximation}
The basic uncertainties of the MSA calculations are given by the
next-to-leading order model corrections such as recoil corrections,
contributions from inelastic double  and triple scattering terms, 
{\it etc.} and  due to the uncertainties of the elementary $I$=0 and 
$I$=1 ${\bar K}N$ scattering lengths.
The calculations of the $K^-\alpha$ scattering length were done for
five sets of parameters for the ${\bar K}N$ lengths shown in
the Table~\ref{Tab1}.
Here we used the results from a $K$-matrix fit (Set 1) and a
separable fit (Set 2)~\cite{Barret}. We also study the constant 
scattering length fit 
(CSL) given by Dalitz and Deloff~\cite{Dalitz}, which we
denoted as Set 3 and the CSL  fit from Conboy~\cite{Conboy} (Set 4).
The recent predictions for ${\bar K}N$ scattering lengths based on the chiral
unitary approach of Ref.\cite{Oller} are denoted as Set~5.

\renewcommand{\arraystretch}{1.2}
\begin{table*}[t]
\begin{center}
\begin{tabular}{|l|c|l|l|l|}
\hline  Set & Reference & $a_0(\bar{K}N) [{\mbox{fm}}]$&
$a_1(\bar{K}N)[{\mbox{fm}}]$& $A(K^- \alpha) [{\mbox{fm}}]$
\\ \hline 
1 & \cite{Barret} & $-1.59+i0.76$ & $0.26 + i .57$ & $-1.80+ i0.90$ \\ \hline 
2 & \cite{Barret} & $-1.61+i0.75$ & $0.32 + i0.70$ & $-1.87 + i 0.95$ \\ \hline 
3 & \cite{Dalitz} & $-1.57+i0.78$ & $0.32 + i0.75$ & $-1.90 + i 0.98$ \\ \hline 
4 & \cite{Conboy} & $-1.03+i0.95$ & $0.94 + i0.72$ & $-2.24+ i 1.58$ \\ \hline
5 & \cite{Oller} & $-1.31+i1.24$ & $0.26 + i0.66$ & $-1.98+ i 1.08$ \\ \hline
\end{tabular}
\end{center}
\caption{\label{Tab1} The $K^- \alpha$ scattering length for various
sets of the elementary $\bar K N $ scattering lengths 
$a({\bar{K}N})$ ($I=0,1$).}
\end{table*}

The results of our calculations are listed in the last
column of Table~\ref{Tab1}. These results are very similar for 
the Sets 1--3 giving  the real and imaginary parts of the
scattering length $A(K^-\alpha)$ within the range of $-1.8 \div -1.9$~fm and 
$0.9 \div 0.98$~fm, respectively. The results for Set 4 are 
quite different: Re$A(K^-\alpha)$ =-2.24~fm and 
Im$A(K^- \alpha)$=1.58 fm. Furthermore, our 
calculations with Set~5 are close to the results obtained with 
Sets 1--3.

Unitarizing the constant scattering length,  we can 
reconstruct the ${\bar K}\alpha$ scattering amplitude within
the zero range approximation (ZRA) as
\begin{equation}
 f_{\bar{K} \alpha}(k)=
\left[A(\bar{K}\alpha)^{-1} - ik\right]^{-1},
\label{f_KHepole} 
\end{equation}
where $k{=}k_{\bar{K}\alpha} $
is the relative momentum of the $K^-\alpha$ system.
The denominator of the amplitude of Eq.(\ref{f_KHepole}) has a zero at 
the complex energy 
\begin{equation}
E^*=E_R - \frac{1}{2}i \Gamma_R=\frac{k^2}{2\mu} ,
\label{pole} 
\end{equation}
where $E_R$ and $\Gamma_R$ are the binding energy and width
of the possible $K^-\alpha$ resonance, respectively. Here
$\mu$ is reduced mass of the system with $\alpha$ mass taken
as 3.728~GeV.

For Set 1 and Set 4 we find a pole at the  complex energies of
$E^*{=}(-6.7{-}i 18/2)$ MeV
and $E^*{=}(-2.0{-}i 11.3/2)$ MeV, respectively. The result
for Set 5 is $E^*{=}(-4.8{-}i 14.9/2)$ MeV.
Note that  assuming a strongly attractive phenomenological 
$\bar K N$ potential, Akaishi and Yamazaki~\cite{Akaishi}
predicted a deeply bound ${\bar K}\alpha$ state at 
$E^*{=}(-86{-}i 34/2)$ MeV, which is far from our solutions. 
With a very similar elementary $\bar K N$ scattering 
length given by Set~1 and used in both calculations, we predict 
a loosely bound state. It is not clear if medium effects and higher
order corrections might be so strong in order to change so drastically
the ${\bar K}\alpha$ scattering length predicted by
our calculations within the  multiple scattering approach. 
In any case it is very important 
to measure the $s$-wave ${\bar K}\alpha$ scattering length 
experimentally and to clarify the situation concerning the 
possible existence of a (deeply) bound ${\bar K}\alpha$ state. 

Note that in the limit of  small absorption,
{\it i.e.} when the imaginary part of $A(\bar K \alpha)$ approaches
zero, the real part of the scattering length should be much larger
for the case of a loosely bound state as compared to the case of
a deeply bound state. Such a situation is supported by the
calculations within ZRA (even in
the presence of absorption) where in
the case of a deeply bound state we found that
$A_{\bar K \alpha}$=$-$0.07{+}$i$0.72 fm.
We expect that the ZRA can be applied for the description of the
amplitude which is generated by the short range potential
used in Ref.\cite{Akaishi}. 
  
\section{ The reaction {\boldmath $dd{\to}\alpha{K^- K^+}$} near 
threshold and the {\boldmath $K^-\alpha$} final-state interaction}

It is well known~\cite{Buescher02,Grishina2001} that the reaction 
\begin{equation}
dd \to \alpha K^- K^+ 
\label{ddHeKK}
\end{equation}
provides an opportunity to study
$I{=}0$ mesonic resonances in the $ K^-K^+$ sector. 

At the same time near the reaction threshold it might be sensitive to 
the to $K^-\alpha $ final state interaction. Here we study 
whether it is possible to evaluate the $s$-wave $K^- \alpha$ 
scattering length from the $K^-\alpha$ final-state interaction.
Similar evaluation of the $d{\bar K^0}$ FSI and relevant scattering length
was done in our previous study~\cite{Sibirtsev04} of  the
$pp{\to}d{\bar K^0}K^+$ reaction. As has been stressed in 
Ref.~\cite{Oset9} this reaction should be very sensitive to the
${\bar K^0}d$ FSI. Through our analysis we extracted a 
new limit for  the $K^-d$ scattering length from the 
$\bar{K^0} d$ invariant mass spectrum from the  
$pp{\to}d{\bar K^0}K^+$ reaction measured recently 
at COSY-J\"ulich~\cite{Kleber}.

It is clear that the FSI effect is essential at low invariant 
masses of the interacting particles, 
where the relative $s$-wave contribution is expected to be dominant. 
One can also safely assume that the range
of the FSI is much larger as compared to the range of the basic hard
interaction related to the production of the $\bar K K$-meson pair.
This means that the basic production amplitude and the FSI term can be 
factorized~\cite{Goldberger,Wycech,Sibirtsev96,Sibirtsev3,Hanhart} and the 
FSI can be taken into account by multiplying the production 
operator by the FSI enhancement factor defined by Eq.(\ref{enhancement}).  

\begin{figure}[tb]
\vspace*{-5mm}
\centerline{\psfig{file=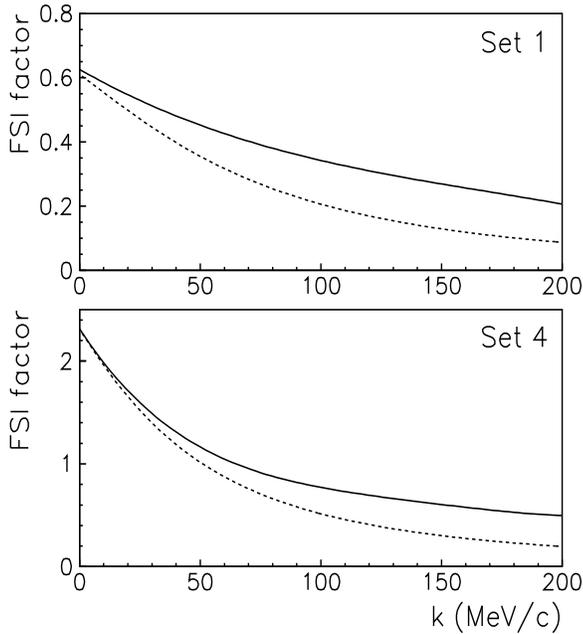,width=8.5cm,height=9.5cm}}
\vspace*{-3mm}
\caption{The $K^- \alpha$ FSI enhancement
factor $\lambda^{\mathrm{MS}}(k)$, Eq.(\ref{enhancement}),
as a function of the relative momentum $k$ of the $K^- \alpha$ system. 
The solid lines in the lower and upper part of the figure
show our calculations with Set~1 and Set~4 for the $\bar K N$ scattering 
lengths, respectively. The dashed lines
illustrate the Watson--Migdal enhancement factor
normalized to $\lambda^{\mathrm{MS}}(k)$ at $k=0$.}
\label{fig:fsi_k}
\end{figure}

Fig.\ref{fig:fsi_k} shows the dependence of the 
$K^-\alpha$ FSI enhancement factor $\lambda^{\mathrm{MS}}(k)$
given by Eq.~(\ref{enhancement})
on the relative momentum of the $K^-\alpha$ system, 
$k$. The solid lines in the upper (lower) part
of Fig.\ref{fig:fsi_k} show the results obtained with  Set 1 (Set 4) 
for the $\bar K N$ scattering length. The calculations with Set 1 
result in $\lambda^{\mathrm{MS}}(k){\simeq}$0.55 at $k{=}0$ and 
FSI  factor smoothly  decreases with $k$. The calculations with Set 4
give $\lambda^{\mathrm{MS}}(k){>}$1 at $k{=}0$ and show a much stronger
$k$-dependence.  

\begin{figure}[tb]
\vspace*{-4mm}
\centerline{\psfig{file=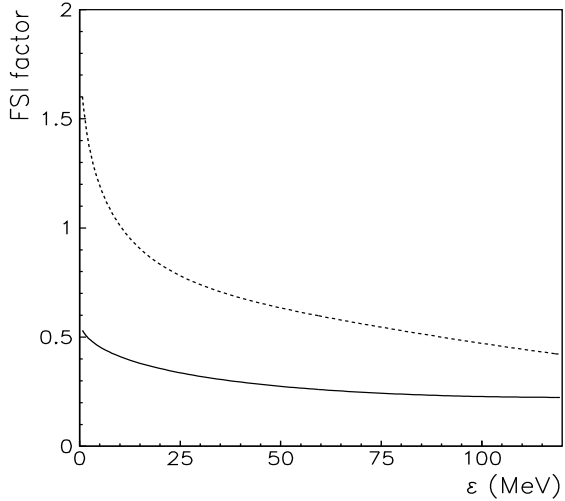,width=8.5cm,height=7.5cm}}
\vspace*{-3mm}
\caption{The $K^-\alpha$ FSI factor averaged over the 
three body phase space of the reaction $dd{\to}\alpha K^+ K^-$
as a function of excess energy. The solid and dashed lines
show the calculations  with parameters of Set~1 and 4, respectively.}
\label{fig2}
\end{figure}

Following the Watson--Migdal approximation~\cite{Watson,Migdal}
the $k$-dependence of the enhancement factor is generally
described in terms of the on-shell scattering amplitude as
\begin{equation}
\lambda_{WM}=\frac{C}{|1-iqA_{\bar K \alpha}|^2},
\end{equation}
where $C$ is normalization constant.

Now, the dashed lines in Fig.~\ref{fig:fsi_k} illustrate the 
Watson--Migdal enhancement factor
normalized to $\lambda^{\mathrm{MS}}(k)$ at $k{=}0$. The upper and
lower parts of Fig.~\ref{fig:fsi_k}    are calculated using 
the scattering lengths $A_{\bar K \alpha}$ obtained with
parameters of Set 1 (Set 4), respectively, and listed in
Table~\ref{Tab1}. It is clear that the  momentum dependence of
$\lambda^{\mathrm{WM}}(k)$ and $\lambda^{\mathrm{MS}}(k)$
is different at different $k$.  However, the absolute difference between
$\lambda^{\mathrm{WM}}(k)$ and  $\lambda^{\mathrm{MS}}(k)$ at
$k{\leq}100$ MeV/c is relatively small.   
 
Obviously, the energy dependence of the total cross section for the  
$dd{\to}\alpha{K^+K^-}$ reaction is also distorted by the the $K^-\alpha$ 
FSI. In Fig.~\ref{fig2} we show the enhancement factor 
$\lambda^{\mathrm{MS}}(k)$ averaged over the 3-body phase space 
as a function of the excess energy $\epsilon$ for
the  $dd{\to}\alpha{K^+K^-}$ reaction. The results for the Sets~2,~3  and 5
are practically the same as for Set~1. It is interesting to note that
there is essentially enhancement of the cross section at small $\epsilon$
for the Set~4, while for the Set~1 we obtain suppression.
The experiment would provide  only a convolution of the
production amplitude and FSI factor. Since the production amplitude
is model dependent it is difficult to extract the absolute
value of the FSI factor from the data. However, the dependence
of the FSI on the relative momentum $k$ is very well
defined because the dependence of the basic hard interaction
on $k$ can be neglected at small $k$. According to 
Ref.\cite{Buescher02} the total cross section of the reaction
$dd{\to}\alpha{K^+K^-}$ might be about 0.4$\ldots$1~nb 
at $\epsilon{=}40\ldots~50$ MeV.   

\begin{figure}[t]
\vspace*{-2mm}
\centerline{\epsfig{file=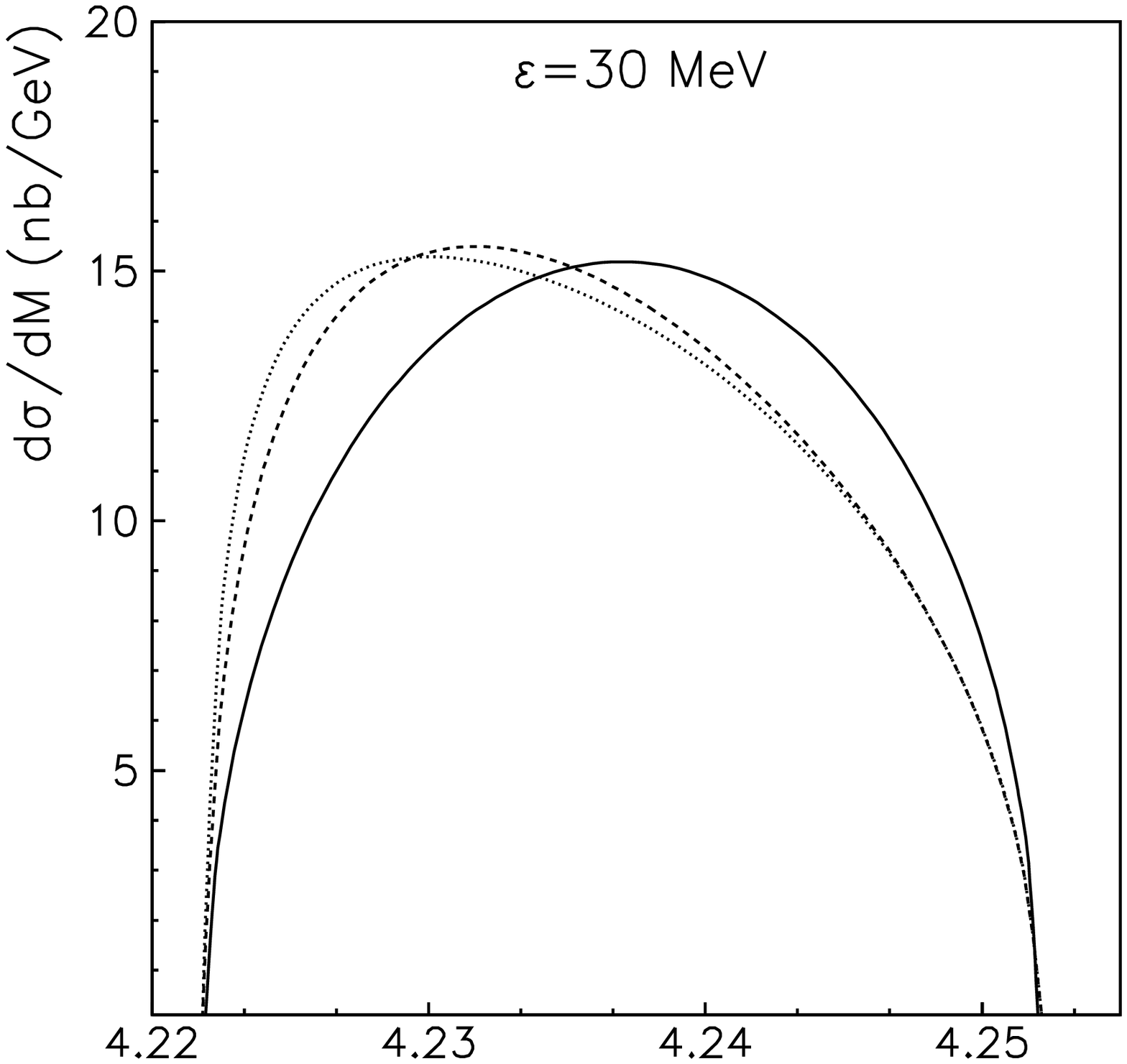,width=8.5cm,height=6.9cm}}
\vspace*{-9mm}
\centerline{\psfig{file=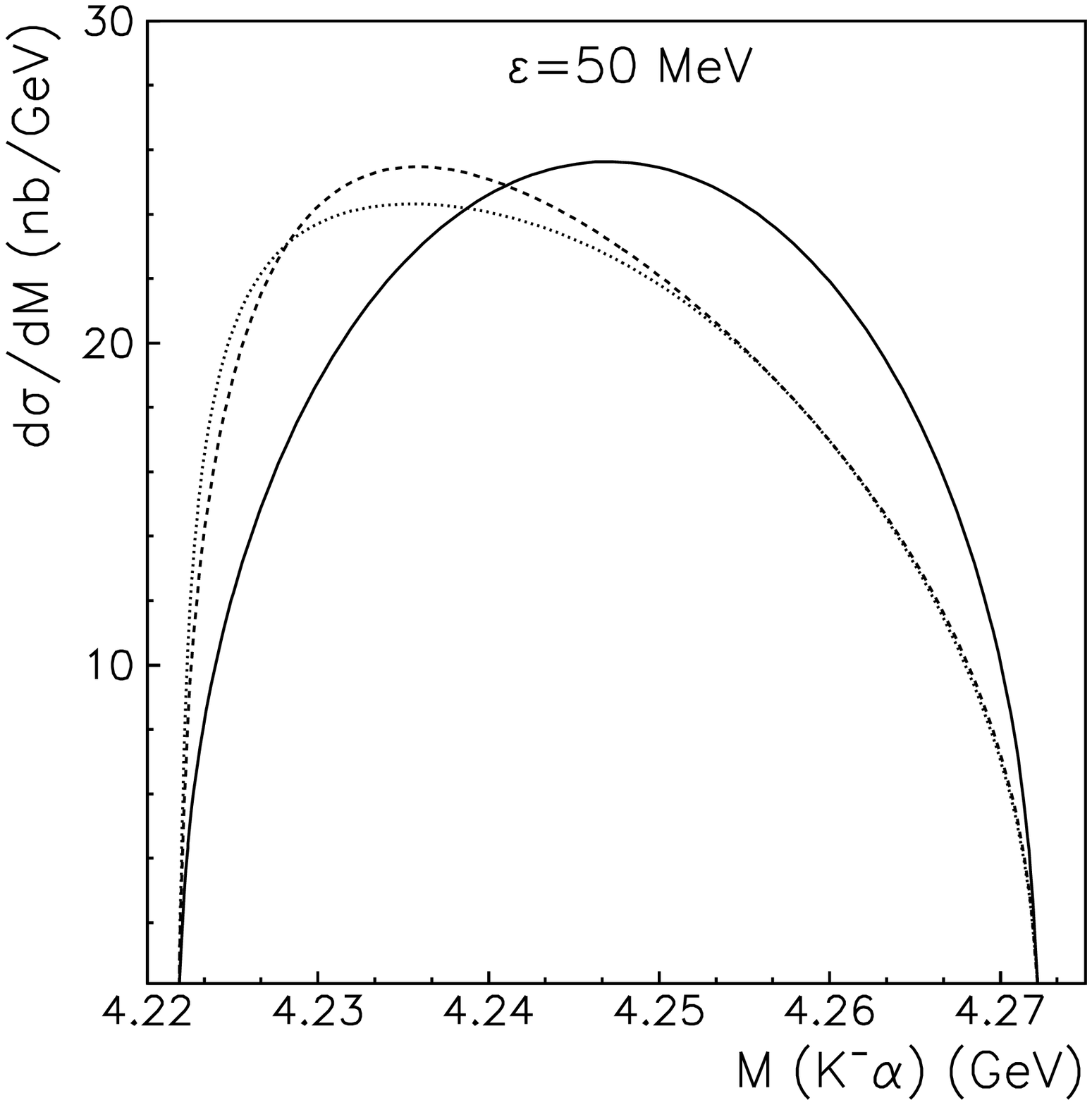,width=8.5cm,height=6.9cm}}
\vspace*{-3mm}
\caption{The invariant $K^-\alpha$ mass spectra produced in the 
$dd{\to}\alpha{K^+ K^-}$ reaction at  excess energies 
30 and 50 MeV.  The solid lines describe the pure phase space 
distribution, while the dashed and dotted lines show our 
calculations with $K^-\alpha$ FSI given by parameters of 
Set 1 and 4, respectively.}
\label{fig3}
\end{figure}

Finally, we calculated the ~$K^-\alpha$ 
invariant mass spectra at excess energies $\epsilon$=30 and $50$~MeV
which are shown in Fig.~\ref{fig3}. The solid lines 
show the calculations for the pure
phase space, {\it i.e.} for the constant production 
amplitude and neglecting FSI.
The dashed and dotted lines in  Fig.~\ref{fig3}  show the results
obtained  with the $K^-\alpha$ FSI calculated with the parameters of
the Set 1 and 4, respectively. All  lines at each figures  are 
normalized to the same value, given by the reaction cross section
at a certain excess energy. At $\epsilon$=50 MeV the invariant mass spectra are
normalized to the   $dd{\to}\alpha{K^+ K^-}$ cross section of 1~nb.  
It is clear that the FSI significantly changes the $K^-\alpha$ mass 
spectra. The most pronounced effect is observed 
at low invariant masses available in the first 10~MeV bin. 

To draw  quantitative conclusions, one can compare the ratio of the
cross sections at the lowest $K^-\alpha$ invariant masses,  within the
first 10~MeV bin, calculated with and without FSI.
We found that this  ratio $R{=}1.26{\ldots}1.34$ at $\epsilon{=}30$ MeV,
$1.49{\ldots}\\ 1.56$ at $\epsilon{=}50$ MeV and $1.84{\ldots}2.18$ at 
$\epsilon{=}100$ MeV.
Here the limits of the ratio at  each excess energy
are given by the calculations with the ${\bar K}N$ scattering
length from  the Set~1 and Set~4. With these estimates it is clear that
reasonable determination of the $K^-\alpha$ scattering length requires
sufficient statistical accuracy at $K^-\alpha$ invariant masses
below 4.23~GeV, at least 100 events. Such a high
precision experiment apparently can be done at COSY. 

\section{Conclusions}
The findings of this study can be summarized as follows:
\begin{itemize}
\item
We have investigated the $s$-wave $K^-\alpha$ scattering length 
and the $K^-\alpha $ FSI enhancement factor within the Foldy--Brueckner 
adiabatic approach based on the multiple scattering  formalism.
We have studied uncertainties of the calculations due to the elementary $K^-N$ 
scattering length available presently. The resulting $s$-wave $K^-\alpha$ 
scattering lengths for the various input parameters are collected in
Tab.~\ref{Tab1}.
\item
Through the determination of the pole position of the 
$K^-\alpha$ scattering amplitude within ZRA, we found a loosely 
bound state with binding energy
$E_R{=}-2{\ldots}-7$MeV and  width $\Gamma_R{=}11{\ldots}18 $~MeV.
Our result differs  from the prediction of Akaishi and 
Yamazaki \cite{Akaishi} obtained under the assumption of a
strongly attractive phenomenological $\bar K N$ potential.
\item
We have analyzed the $K^-\alpha$ FSI in the reaction  
$dd{\to}\alpha K^+ K^-$ and discussed the possibility to evaluate the
$K^-\alpha$ scattering length from the  $K^-\alpha$ invariant mass spectra.
We have demonstrated that the measurement of the 
$K^-\alpha$ mass distribution  near the reaction threshold may 
provide a new tool for the determination of the $s$-wave 
$K^-\alpha$ scattering length.
\item
Furthermore, we have investigated the momentum dependence of the enhancement 
factor $\lambda^{\mathrm{MS}}(k)$ calculated within MSA and compared it 
with the one obtained utilizing the Watson--Migdal formalism. It was found that
the absolute difference between both calculations is relatively small
at momenta $q{\leq}100$ MeV/c.   
\end{itemize}

It is important to stress that for  kaonic helium atoms, energy shifts 
can be measured for the $2p$ state and widths for the $2p$ and $3d$ states. 
The $np{\to}1s$ transitions for $^4$He cannot be observed since the
absorption from the $p$ states is almost complete~\cite{Batty}.
Therefore the possibility to determine the $s$-wave $\bar K \alpha $ 
scattering length from experiments with kaonic atoms is questionable.
With this respect a measurement at COSY provides an unique opportunity
to determine  $s$-wave  $K^-\alpha$ scattering length.

\subsection*{Acknowledgements}
We appreciate discussions  with
C.~Hanhart, M.~Hartmann, R.~Lemmer and P.~Winter. 
This work was partially supported by Deutsche Forschungsgemeinschaft 
through funds provided to the SFB/TR 16 ``Subnuclear Structure of Matter''
and  by the DFG grant 436 RUS   113/787.
This research is part of the EU Integrated Infrastructure Initiative Hadron
Physics Project under contract number RII3-CT-2004-506078.
V.G. acknowledges support by the COSY FFE grant No. 41520739 and A.S.
acknowledges support by the COSY FFE grant No. 41445400 (COSY-067).
% Author, Journal \textbf{Volume}, (year) page numbers.
% Author, \textit{Book title} (Publisher, place year) page numbers

\end{document}